\documentclass[preprint,amssymb,amsmath]{article}

\usepackage{graphicx}
\usepackage{pifont}
\usepackage{textcomp}
\graphicspath{{fig/}}
\usepackage{dcolumn}
\usepackage{bm}
\usepackage{float}
\usepackage{amssymb}
\usepackage{amsmath}
\usepackage{color}
\begin{document}

\title{ Bose-Einstein condensation of magnons \\
in spin pumping systems  }

\author{Kouki Nakata and Yusuke Korai    \\    \   \\
 Yukawa Institute for Theoretical Physics, 
Kyoto University,  \\
Kitashirakawa Oiwake-Cho, Kyoto 606-8502, Japan,  \\
nakata@yukawa.kyoto-u.ac.jp  \\
korai@yukawa.kyoto-u.ac.jp
}

\maketitle

\begin{abstract}
We clarify the condition for the occurrence of magnon Bose-Einstein condensation (BEC)  in spin pumping systems
without using external pumping magnetic fields.
The Goldstone model is generalized and the stability of the vacuum is closely investigated.
By applying the generalized Goldstone model to  spin pumping systems,
the condition for the experimental realization of the stable magnon BEC in  spin pumping systems
is theoretically proposed.
\end{abstract}

\maketitle

\section{Introduction}
\label{sec:intro}

The experimental observation of  BEC\cite{leggett} in variety kinds of systems  
as well as trapped ultracold atoms and molecules have recently been reported;
photons in an optical microcavity,\cite{klaers}
semiconductor microcavity exciton polaritons,\cite{deng}
and microcavity polaritons in a trap\cite{balili} et al.\cite{kasprzak} 
This fact implies the universal aspects  of this phenomenon.
BEC is, in principle, the phenomenon that a macroscopic number of particles occupies a single-particle state.\cite{leggett,masahito}
Thus  quasiparticles also undergo BEC  and in particular,
magnon BEC\cite{demokritov,bunkov} has now become one of the most attractive subjects
in condensed matter physics.

\begin{figure}[h]
\begin{center}
\includegraphics[width=5cm,clip]{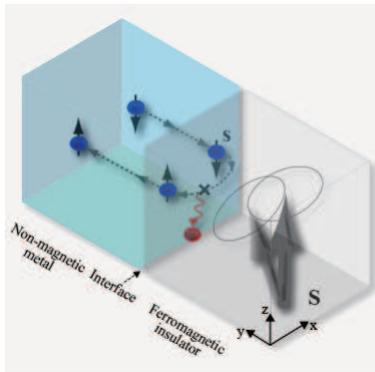}
\caption{(Color online).
The schematic picture of   spin pumping systems;
spheres  represent magnons and those with arrows are conduction electrons.
The interface is characterized by the exchange interaction between conduction electrons and the ferromagnet 
$  {V}_{\rm{ex}}$.
Thus, the interface is defined as an effective area where the Fermi gas (i.e. conduction electrons)  and the Bose gas (i.e. magnons) coexist to interact.
Conduction electrons cannot enter the ferromagnet, which is an insulator.
{\textbf{\textit{Clear pictures are available at the following URL;}}}
https://dl.dropbox.com/u/5407955/MagnonBECinSP.pdf
 \label{fig:pumping} }
\end{center}
\end{figure}

In this paper,
we go after the possibility for the occurrence of magnon BEC\cite{bunkovalone,bunkov} 
without using external pumping magnetic fields (i.e. quantum fluctuations)\cite{QSP,demokritov}
in spin pumping systems (Fig. \ref{fig:pumping}).
At the interface of a ferromagnetic insulator and non-magnetic metal junction, 
conduction electrons $\mathbf{s} $  interact with ferromagnetic localized spins  $\mathbf{S} $;
$V_{\rm{ex}}=  -  \mathbf{S} \cdot  \mathbf{s} $.
The degree of freedom of ferromagnetic localized spins are reduced to that of magnons\cite{spinwave} via the Holstein-Primakoff transformation.
Therefore we regard the interface of  spin pumping systems as the effective area 
where magnons interact with conduction electrons;\cite{QSP,TSP} 
the interface  can be regarded as a ferromagnetic metal.\cite{bauer}
The exchange interaction $  {V}_{\rm{ex}}$ at the interface is essential to spin pumping\cite{saitohprivate}
and hence, we identify  the system characterized $  {V}_{\rm{ex}}$  with the spin pumping system.
From now on, we exclusively focus on the dynamics at the interface  (Fig. \ref{fig:pumping}).
To clarify the condition for the experimental realization of the stable magnon BEC state at the interface of the spin pumping system
is the final goal of this paper.

Originally, spin pumping systems have been attaching special attention from the viewpoint of spintronics,
which is a rapidly developing new branch of physics.
The central theme is the active manipulation of spin degrees of freedom as well as charge ones of electrons.
Thus by going after the possibility for the occurrence of magnon BEC in spin pumping systems,
we   build a bridge between the research on spintronics\cite{uchidaTSP,mod2} and  magnon BEC.\cite{oshikawa,ueda}

In this paper,
we employ a non-perturbative theory to go beyond the perturbative analysis\cite{TSP}
and investigate the possibility for the occurrence of magnon BEC in  spin pumping systems
without using external pumping magnetic field.\cite{demokritov,QSP}
For the purpose,
we generalize the Goldstone model in sec. \ref{sec:SSB}
and apply the generalized Goldstone model to spin pumping systems in sec. \ref{sec:magnonBEC}. 
To provide details, this paper is structured as follows;
first, 
by adopting the usual Goldstone model\cite{peskin,coleman} as an example,
we quickly review the idea of spontaneous symmetry breaking (SSB) in the classical field theory in sec. \ref{subsec:Goldstone}.
Second, 
in sec. \ref{subsec:gene}, by introducing a new complex scalar field with keeping the $U(1)$-symmetry of the system,
we minimally generalize the above standard Goldstone model so as to include the effects of other degrees of freedoms
(e.g. spins carried by conduction electrons).
The condition for the occurrence of the $U(1)$-SSB is clarified in the minimally generalized model.
On top of this, the stability of the vacuum is  closely investigated.
Last,  
by applying the above minimally generalized Goldstone model to anisotropic  spin pumping systems,
we go after the possibility for the occurrence of magnon BEC in sec. \ref{subsec:aniso}.
By further extending the Goldstone model, 
the condition for the experimental realization of the stable magnon BEC state in  spin pumping systems is theoretically proposed
in sec. \ref{subsec:geneGoldstone}.
This is the main aim of this paper.

\section{BEC and SSB}
\label{sec:SSB}

The experimental realizations of magnon BEC in variety kinds of materials have been reported;
TlCuCl$_3$,\cite{oshikawa} 
Cs$_2$CuCl$_4$,\cite{radu} 
Yttrium-iron-garnet (YIG),\cite{demokritov,demidov,chumak} 
and BaCuSi$_2$O$_6$\cite{ruegg} et al.
In the present study,
we identify the expectation value of the bosonic annihilation operator $\langle \Psi  \rangle$ with 
the macroscopic condensate order parameter\cite{oshikawa}
and adopt  as the criterion for the occurrence of BEC.\cite{bunkov,totsuka}
That is, a non-zero value of the order parameter $ \langle \Psi  \rangle \not=0$ under the $U(1) $-symmetric Hamiltonian
does mean the occurrence of BEC, which is accompanied by $U(1) $-SSB.
This definition of BEC has now been very commonly used in the literature.\cite{bunkov,leggett,totsuka}

On the basis of this definition of BEC,
we investigate the possibility for the occurrence of the $U(1)$-SSB of the vacuum in spin pumping systems,
which is accompanied by  a non-zero value of the order parameter under the $U(1)$-symmetric Hamiltonian.
In order to go beyond  the perturbative analysis by the Schwinger-Keldysh formalism,\cite{TSP}
we employ a powerful theoretical technique `non-perturbative theory',\cite{peskin}  which does not rely on the assumption called the adiabatic theorem
(i.e. the well-known Gell-Mann and Low theorem).\cite{fetter,peskin,Gell-mann}
Therefore we can analyze beyond a perturbative theory.\cite{TSP}

\begin{figure}[h]
\begin{center}
\includegraphics[width=8cm,clip]{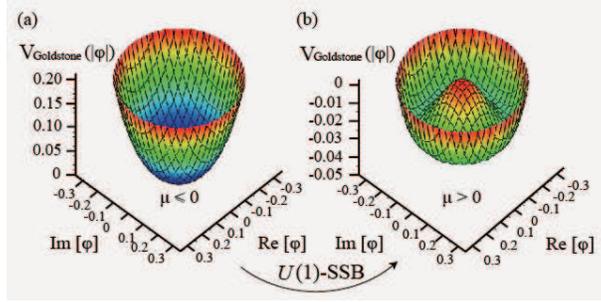}
\caption{(Color online).
Schematic pictures of the $U(1)$-SSB of the vacuum
in the Goldstone model $  V_{\rm{Goldstone}} (\mid \varphi \mid )$; eq. (\ref{eqn:ssb5}).
When the chemical potential becomes positive ($0<\mu$),
the Goldstone model forms `the Mexican-hat potential'\cite{amore} (b) and 
the $U(1)$-symmetry of the vacuum is spontaneously broken; eq. (\ref{eqn:ssb6}).
As an example,  each parameter is set as follows;
(a) $ {\mathcal{J}} = +10$, $ \mu =-1 $ and
(b) $ {\mathcal{J}} = +10$, $ \mu =+1 $.
It will be useful to see also Fig. \ref{fig:SSBfeeble} (a). 
 \label{fig:SSBstabdard} }
\end{center}
\end{figure}

\subsection{Goldstone model}
\label{subsec:Goldstone}

Before going on to the main subject,
let us briefly review\cite{peskin,coleman,altland} the idea of the SSB of the vacuum in classical field theory.
As an example, 
we employ the (so called) `Goldstone model' whose potential term is given as  (see also Fig. \ref{fig:SSBstabdard})
\begin{subequations}
\begin{eqnarray}
    V_{\rm{Goldstone}}(\varphi ) &:=& -\mu  \varphi \varphi ^{\ast }        +{\mathcal{J}}   (\varphi \varphi ^{\ast }  )^2     \\
                                                            &= &  -\mu   \mid  \varphi    \mid ^2    +\mathcal{J}  \mid  \varphi    \mid ^4,    
\label{eqn:ssb1}  
\end{eqnarray}     
\end{subequations}
in which  the variable $ \varphi ( \in  {\mathbb{C}})$ denotes a complex scalar field.
The parameter  $\mu (\in  {\mathbb{R}})$ represents the dimensionless chemical potential
and $ {\mathcal{J}} (\in  {\mathbb{R}})$ does a dimensionless coupling constant. 
It is clear that the Goldstone model possesses the global $U(1)$-symmetry;
 $  \varphi \mapsto  {\rm{e}}^{i\theta }\varphi     \    {\rm{with}}   \       \theta \equiv ({\rm{const.}})   \in  {\mathbb{R}}$. 
For the stability of the system or the vacuum (i.e. the ground state),
there should be a lower bound on the energy level of the system.\cite{Affleck}
Thus  the condition is required (see also Fig. \ref{fig:SSBfeeble} (a));
\begin{eqnarray}
      0<     \mathcal{J}.
\label{eqn:ssb4}  
\end{eqnarray}            

From here on,
we will assume that the possible vacuum states are invariant under translations and 
they are time-independent.\cite{peskin,coleman} 
Thus,  the candidate of the stable vacuum of the system is given 
as the stationary point of the effective potential  $   V_{\rm{Goldstone}}^{\rm{eff}} $;\cite{peskin,altland}
\begin{eqnarray}
    \frac{\partial  V_{\rm{Goldstone}}^{\rm{eff}} }{\partial  \varphi } = 0.
\label{eqn:ssb3-2}   
\end{eqnarray}            
In addition,
within the classical theory  in the sense that 
we omit the quantum effects (i.e. loop corrections)
and discuss within the tree-level,
the effective potential $V_{\rm{Goldstone}}^{\rm{eff}}$ is reduced to 
the usual one $V_{\rm{Goldstone}}$;\cite{peskin}
\begin{subequations}
\begin{eqnarray}
    V_{\rm{Goldstone}}^{\rm{eff}} (\mid \varphi \mid )
&=&  -\mu   \mid  \varphi    \mid ^2    +\mathcal{J}  \mid  \varphi    \mid ^4    + {\cal{O}}(\hbar )   \\
 &=&     V_{\rm{Goldstone}}(\mid \varphi \mid )        + {\cal{O}}(\hbar ).            
\label{eqn:ssb3}   
\end{eqnarray}     
\end{subequations}       
Thus the minimum-energy classical configuration is a uniform field $ \varphi  = \varphi _0 $
with $\varphi _0 $ chosen to minimize the  potential $   V_{\rm{Goldstone}} $;\cite{peskin,coleman}
\begin{eqnarray}
    V_{\rm{Goldstone}} (\mid \varphi \mid ) = \mathcal{J} \Big[    \mid \varphi \mid ^2   -\frac{\mu}{2 \mathcal{J}} \Big]^2  - \frac{{\mu}^2}{4 \mathcal{J} }.
\label{eqn:ssb5}  
\end{eqnarray}            
Consequently   when  the chemical potential is positive ($  0<  \mu $),  
the vacuum expectation value of  the field  $ \varphi_0 $  reads 
\begin{eqnarray}
     \mid \varphi_0 \mid &=& \sqrt{\frac{\mu }{2 \mathcal{J} }}    \   \    (\not=0). 
\label{eqn:ssb6}      
\end{eqnarray}    
As the result,  the  $U(1)$-SSB  of the vacuum  does occur
when  the chemical potential is positive ($  0<  \mu $, Fig. \ref{fig:SSBstabdard} (b));
otherwise not (i.e. $\mu \leq 0  $, Fig. \ref{fig:SSBstabdard} (a)).

\subsection{Minimally generalized Goldstone model}
\label{subsec:gene}

We have seen that the  $U(1)$-symmetry  of the vacuum  in the Goldstone model 
where only one complex scalar field $\varphi $ acts 
is spontaneously  broken
when the chemical  potential $\mu$ is properly adjusted;
this is the rigorous theoretical result based on the non-perturbative analysis.
The Goldstone model has been used to describe a dilute Bose gas 
in the classical limit at $T=0$ as a (phenomenological) standard model;\cite{altland,amore}
e.g. magnons, regardless\cite{totsuka} of ferromagnets\cite{tupitsyn,bunkov} or antiferromagnets.\cite{oshikawa,ueda,amore,gia}
On the other hand, in real materials and experiments,
there does exist variety kinds of freedoms  besides the one on which we focus,
such as magnetic impurities, phonons,\cite{hick} and photons\cite{demokritov} et al.  
(see also sec. \ref{subsec:aniso}).
Therefore it is desirable to extend the Goldstone model so as to 
include the effects of such degrees of freedoms
by introducing a new complex scalar field $\psi (\in  {\mathbb{C}})$
which couples with the usual field $\varphi $.

Now, our strategy of the generalization of the Goldstone model reads as follows;
for clearness,
we exclusively focus on  when `the  usual chemical potential $\mu$'    is negative ($ \mu \leq 0  $). 
In that case, the model reads
\begin{eqnarray}
    V_{\rm{Goldstone}}(\varphi ) &=& {\cal{B}}  \mid  \varphi    \mid ^2    +  \mathcal{J}  \mid  \varphi    \mid ^4,
\label{eqn:ssb7}      
\end{eqnarray}            
in which we have denoted as, $  - \mu  =:  {\cal{B}}    \   (\geq 0)$, for convenience (see also sec. \ref{subsec:aniso}).\footnote{
Note that the sign of ${\cal{B}}$ is opposite from the one of  the chemical potential $\mu$.
}
In this case,
it is apparent that the vacuum expectation value of  the field becomes zero  ($ \varphi_0=0 $),
which is not accompanied by  the $U(1)$-SSB of the vacuum (Fig. \ref{fig:SSBstabdard} (a)).
Now, we go after the possibility for the occurrence of the $U(1)$-SSB of the vacuum
owing to the coupling with other degrees of freedom represented by a complex scalar field $\psi (\in  {\mathbb{C}})$ such as 
\begin{subequations}
\begin{eqnarray}
      &  (\varphi \psi ^{\ast }  + \varphi ^{\ast }\psi  ),  \    \    \psi \psi ^{\ast },&             \\                               
                  &\rm{and}&    \nonumber  \\
                    &    \mid \varphi \mid ^2 \mid \psi \mid ^2,     &                                   
\label{eqn:ssb9}     
\end{eqnarray} 
\end{subequations}           
which do not violate the $U(1)$-symmetry of the system;
                  $     (\varphi ,\psi )\mapsto   {\rm{e}}^{i\theta }(\varphi ,\psi ),   \   \     {\rm{with}}    \  \   \theta\equiv (\rm{const.}) \in {\mathbb{R}}$.                                             
It is expected that these couplings bring   `effective chemical potential' to $\varphi  \  (\varphi ^*)$.
As the result, the total chemical potential might  become positive 
and the $U(1)$-SSB  of the vacuum might be generated.

\subsubsection{Minimal model}
\label{subsubsec:mini}

We introduce the minimally generalized Goldstone model $V_{  U(1){\mathchar`-}{\rm{mini.}}} (\varphi , \psi )$ 
by adding couplings,  $    (\varphi \psi ^{\ast }  + \varphi ^{\ast }\psi  ) $ and  $ \psi \psi ^{\ast }$,   into the Goldstone model;
\begin{subequations}
\begin{eqnarray}
    V_{  U(1){\mathchar`-}{\rm{mini.}}} (\varphi , \psi )&:=&  V_{\rm{Goldstone}}(\varphi )  
                                                       - \gamma  (\varphi \psi ^{\ast }  + \varphi ^{\ast }\psi  ) +\kappa \psi \psi ^{\ast }     \\
                                                        &=&     {\mathcal{B}} \mid \varphi  \mid ^2  + {\mathcal{J}}  \mid \varphi   \mid ^4    
                                                                      -\gamma  (\varphi \psi ^{\ast }  + \varphi ^{\ast }\psi  ) +\kappa   \mid  \psi  \mid ^2,
\label{eqn:ssb12}      
\end{eqnarray}  
\end{subequations}          
where each variable,   $ \gamma  (\in {\mathbb{R}}  \   {\rm{and}}  \  \gamma >0) $  and  $\kappa (\in {\mathbb{R}})$, represents  a dimensionless coupling constant.
The minimally generalized Goldstone model $ V_{  U(1){\mathchar`-}{\rm{mini.}}}$  includes two kinds of fields, $ \varphi $ and $ \psi $.
Therefore for the stability of the vacuum, there should be a lower bound on the energy level of the system 
in respect to $ \psi $ as well as $\varphi $;\cite{totsuka}
in terms of $\varphi $,
the minimally generalized Goldstone model can be expressed as
\begin{eqnarray}
 V_{  U(1){\mathchar`-}{\rm{mini.}}} (\varphi )&=&  {\mathcal{J}}  \mid \varphi   \mid ^4+  {\mathcal{B}} \mid \varphi  \mid ^2   
                                                       -\gamma  (\varphi \psi ^{\ast }  + \varphi ^{\ast }\psi  )   
                                                       +  {\cal{O}}  ( (\varphi  ^{(\ast )})^0 ).    
\label{eqn:ssb01}      
\end{eqnarray}            
Therefore the condition is required; $ 0< {\mathcal{J}} $.
In addition,   from the viewpoint of  $\psi$,
$ V_{  U(1){\mathchar`-}{\rm{mini.}}}$ can be regarded  as
\begin{eqnarray}
   V_{  U(1){\mathchar`-}{\rm{mini.}}} ( \psi ) &=&  \kappa   \mid  \psi  \mid ^2   -  \gamma  (\varphi \psi ^{\ast }  + \varphi ^{\ast }\psi  )   
+ {\cal{O}}  ( (\psi ^{(\ast )})^0 ). 
\label{eqn:ssb02}      
\end{eqnarray}            
Thus the condition should be satisfied;  
\begin{eqnarray}
  0<\kappa. 
\label{eqn:ssb13}  
\end{eqnarray}            
Otherwise, 
`the saddle point'  (see  Fig. \ref{fig:SSBstability} (a) as an example)   cannot be eliminated
from the condition for the stationary point represented by eq. (\ref{eqn:ssb3-2});
the saddle point gives an unstable state and 
the situation is out of the aim of the present study.

\begin{figure}[h]
\begin{center}
\includegraphics[width=6cm,clip]{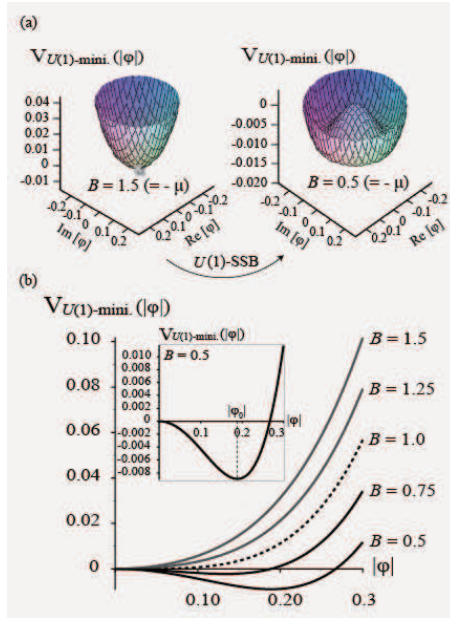}
\caption{(Color online).
Schematic pictures of the $U(1)$-SSB of the vacuum
in the minimally generalized Goldstone model $  V_{U(1){\mathchar`-}{\rm{mini.}}} (\mid  \varphi \mid  )$; eq. (\ref{eqn:ssb15}).
 As an example,  each dimensionless parameter is set as follows;
$ {\mathcal{J}}  =7$ and   $  \kappa = \gamma  =1$. 
Therefore for the occurrence of the  $U(1)$-SSB,
the parameter $ {\mathcal{B}}$ must satisfy the condition (inequality (\ref{eqn:ssb16})); $ (0 <) {\mathcal{B}}  <1$.
(a) 
Even when parameter $\mu$ is negative ($0 < {\cal{B}} := -\mu $, see also Fig. \ref{fig:SSBstabdard} (a)),
the $U(1)$-SSB of the vacuum can occur in the minimally generalized Goldstone model 
because `the effective potential'  is not $\mu$, but $  ( {\gamma  ^2}/{\kappa } - {\mathcal{B}}) $ (eq. (\ref{eqn:ssb15a})).
(b) 
When $ {\mathcal{B}}  =0.5$,
the $U(1)$-symmetry of the vacuum is spontaneously broken and 
the vacuum expectation value $\varphi _0$ (eq. (\ref{eqn:ssb17}))  becomes
 $ \mid  \varphi _0 \mid \simeq  0.189$.
 \label{fig:SSBU1} }
\end{center}
\end{figure}

Under these conditions (i.e. inequalities (\ref{eqn:ssb4}) and (\ref{eqn:ssb13})), 
through the same procedure with sec. \ref{subsec:Goldstone} and within the classical theory,
we seek the true stable vacuum of the minimally generalized Goldstone model.
The condition for the stationary point in respect to $\psi $  gives
\begin{eqnarray}
   \frac{\partial V_{U(1){\mathchar`-}{\rm{mini.}}}}{\partial \psi }   =  0     \Rightarrow 
    \psi ^{\ast }  = \frac{\gamma  }{\kappa } \varphi ^{\ast }.
\label{eqn:ssb14}    
\end{eqnarray}         
On the point,
$ V_{U(1){\mathchar`-}{\rm{mini.}}}  $  can be rewritten as 
\begin{subequations}
\begin{eqnarray}
    V_{U(1){\mathchar`-}{\rm{mini.}}} (\varphi , \psi = \frac{\gamma  }{\kappa }\varphi )&=&  
({\mathcal{B}}-\frac{\gamma  ^2}{\kappa }) \mid  \varphi  \mid ^2  +{\mathcal{J}}\mid \varphi   \mid ^4  \label{eqn:ssb15a}        \\
&=& {\mathcal{J}}\Big[   \mid  \varphi  \mid ^2   - \frac{1}{2{\mathcal{J}}} (\frac{\gamma  ^2}{\kappa } -{\mathcal{B}})  \Big]^2     
-\frac{1}{4{\mathcal{J}}} (\frac{\gamma  ^2}{\kappa }-{\mathcal{B}})^2     \\
&=:& V_{U(1){\mathchar`-}{\rm{mini.}}} (\mid  \varphi \mid  ).
\label{eqn:ssb15}     
\end{eqnarray}  
\end{subequations}          
It is clear that
the minimally generalized Goldstone model  $V_{U(1){\mathchar`-}{\rm{mini.}}} $ is reduced to the standard one $ V_{\rm{Goldstone}} $
with `the effective potential' $  ( {\gamma  ^2}/{\kappa } - {\mathcal{B}}) $; as expected,
the coupling with other degrees of freedoms $ \psi $  has brought
`the effective chemical potential' ${\gamma  ^2}/{\kappa }$ (see eq. (\ref{eqn:ssb15a})).
As the result, the total chemical potential can become positive and   hence,
the $U(1)$-SSB   of the vacuum  occurs (see Fig. \ref{fig:SSBU1} (a))  when 
\begin{eqnarray}
   (0<)  \   {\mathcal{B}}  < \frac{\gamma  ^2}{\kappa }.
\label{eqn:ssb16}   
\end{eqnarray}            
Under this condition,
the vacuum expectation value $\varphi _0 $  (see Fig. \ref{fig:SSBU1} (b))  reads
\begin{eqnarray}
              \mid \varphi _0\mid &=&\sqrt{ \frac{1}{2{\mathcal{J}}} (\frac{\gamma  ^2}{\kappa }-{\mathcal{B}})}.
\label{eqn:ssb17}     
\end{eqnarray}         

Here let us emphasize that 
when $ \gamma  =0 $ or $ \kappa=0 $ (see inequalities (\ref{eqn:ssb13}) and (\ref{eqn:ssb16})),
the $ U(1) $-SSB of the vacuum  cannot occur.
That is, the $\gamma $-term as well as the  $\kappa$-term  in eq. (\ref{eqn:ssb12}) is essential 
for the occurrence of the $ U(1) $-SSB  of the vacuum.
Therefore we have named $V_{U(1){\mathchar`-}{\rm{mini.}}}$
the `minimally' generalized Goldstone model.

\subsubsection{Stability of  vacuum}
\label{subsubsec:stability}

\begin{figure}[h]
\begin{center}
\includegraphics[width=7cm,clip]{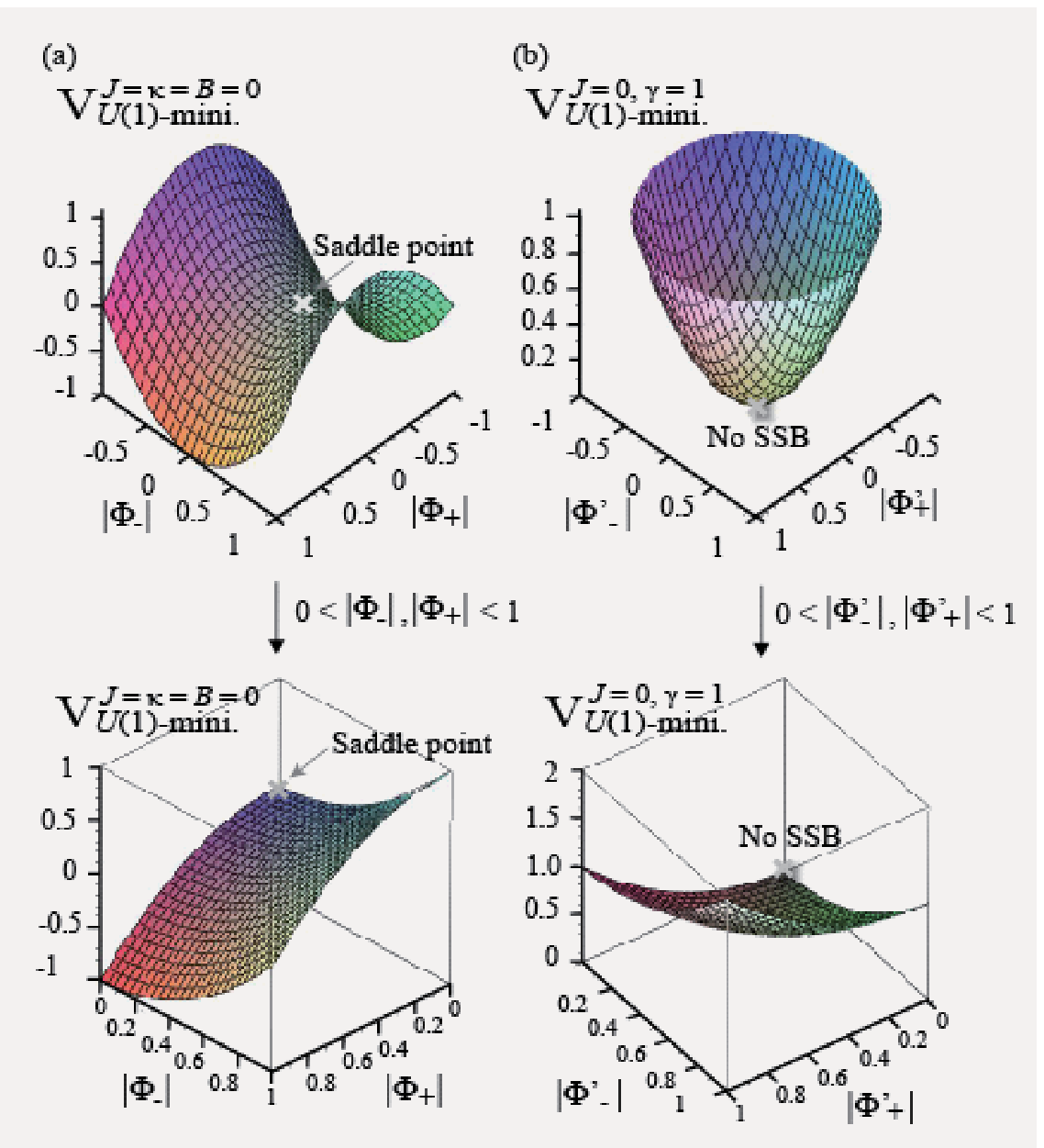}
\caption{(Color online).
(a)  
Plot of  $V_{U(1){\mathchar`-}{\rm{mini.}}}^{{\mathcal{J}}=\kappa ={\mathcal{B}}=0}$ (eq. (\ref{eqn:ssb27-2})) with  $ \gamma  =1 $.
The origin (i.e. $  \mid  \Phi _+ \mid =\mid \Phi _-\mid =0  $) is a saddle point, which is not stable or the true vacuum;
there are no true stable vacuum states in $ V_{U(1){\mathchar`-}{\rm{mini.}}}^{{\mathcal{J}}=\kappa ={\mathcal{B}}=0}$.
(b) Plot of  $V_{U(1){\mathchar`-}{\rm{mini.}}}^{{\mathcal{J}}=0, \gamma  =1}$ (eq. (\ref{eqn:ssb34})) with $ \lambda _{+} = \lambda _{-} = +1$.
The origin (i.e. $  \mid  \Phi^{\prime} _+ \mid =\mid \Phi^{\prime} _-\mid =0  $)  is the original vacuum;
the state is not generated by the $U(1)$-SSB
and it in fact gives $ \varphi _0 =0$. 
(a) (b)
Note that  although  the variables are restricted to $ 0 \leq  \mid  \Phi _{\pm }^{(\prime)} \mid  $ by definition,
we also have plotted the region; $ -1 \leq  \mid  \Phi _{\pm }^{(\prime)} \mid  $ for clearness.
 \label{fig:SSBstability} }
\end{center}
\end{figure}

It would be useful to investigate the stability of the vacuum 
to confirm the importance of the repulsive interaction,\cite{Affleck} 
$ {\mathcal{J}} \mid \varphi \mid ^4 $ with $  0 <  {\mathcal{J}} $,
for the realization of the stable vacuum.
For simplicity here,
each coupling  constant  in eq. (\ref{eqn:ssb12})  is set as follows;
$  {\mathcal{J}}=\kappa ={\mathcal{B}}=0  $.
On this condition, 
$  V_{U(1){\mathchar`-}{\rm{mini.}}}  $ becomes
\begin{subequations}
\begin{eqnarray}
    V_{U(1){\mathchar`-}{\rm{mini.}}}^{{\mathcal{J}}=\kappa ={\mathcal{B}}=0} (\varphi , \psi  )&=& 
    -\gamma  (\varphi \psi ^{\ast }  + \varphi ^{\ast }\psi  )   \\
 &=& -\gamma  
\begin{pmatrix}
  \varphi ^{\ast }   &   \psi ^{\ast }   \\    
\end{pmatrix}
{\mathcal{A}}
\begin{pmatrix}
     \varphi    \\   \psi    
\end{pmatrix},
\label{eqn:ssb24}   
\end{eqnarray}      
\end{subequations}      
with
\begin{equation}
{\mathcal{A}}:=
\begin{pmatrix}
  0   &   1   \\   1   &  0   
\end{pmatrix}
= {\mathcal{A}}^{\dagger }.
\label{eqn:ssb25}   
\end{equation}
It is clear that 
$  V_{U(1){\mathchar`-}{\rm{mini.}}}^{{\mathcal{J}}=\kappa ={\mathcal{B}}=0} $  takes quadratic form
and the matrix $\mathcal{A}$ is Hermitian.
Therefore  $  V_{U(1){\mathchar`-}{\rm{mini.}}}^{{\mathcal{J}}=\kappa ={\mathcal{B}}=0} $
can be easily diagonalized\cite{totsuka} 
via an unitary matrix $ U$ as follows (see also Appendix \ref{sec:diagonalization});
\begin{eqnarray}
    V_{U(1){\mathchar`-}{\rm{mini.}}}^{{\mathcal{J}}=\kappa ={\mathcal{B}}=0} 
 (\mid  \Phi_{+}\mid ,   \mid \Phi_{-}\mid  )
  &=&  -\gamma  (  \mid  \Phi_{+}   \mid^{2}   -  \mid   \Phi_{-}   \mid^{2}   ),   \label{eqn:ssb27-2}   
\end{eqnarray}            
with
\begin{eqnarray}
  \frac{1}{\sqrt{2}}
\begin{pmatrix}
 \varphi + \psi     \\      \varphi    -   \psi        
\end{pmatrix} 
&=:& \begin{pmatrix}
 \Phi _+     \\     \Phi _-       
\end{pmatrix}. 
\label{eqn:ssb27}   
\end{eqnarray}            
Fig. \ref{fig:SSBstability} (a) describes 
$ V_{U(1){\mathchar`-}{\rm{mini.}}}^{{\mathcal{J}}=\kappa ={\mathcal{B}}=0}  ( \mid \Phi_{+}\mid  ,   \mid \Phi_{-} \mid ) $.
It is apparent that the origin (i.e. $  \mid  \Phi _+ \mid =\mid \Phi _-\mid =0  $) has become `a saddle point',
which is not stable or the true vacuum;
there are no true stable vacuum states in $ V_{U(1){\mathchar`-}{\rm{mini.}}}^{{\mathcal{J}}=\kappa ={\mathcal{B}}=0}$.

Note that even when a non-zero value is given to each coupling constant ($ \kappa $ and  $ \mathcal{B}$),
the situation does not change
 as long as $ {\mathcal{J}}$ is zero;
for simplicity here, we take $ \gamma  =1$.
On this condition,
$  V_{U(1){\mathchar`-}{\rm{mini.}}}  $ becomes (see also Appendix \ref{sec:diagonalization})
\begin{subequations}
\begin{eqnarray}
    V_{U(1){\mathchar`-}{\rm{mini.}}}^{{\mathcal{J}}=0, \gamma  = 1} (\varphi , \psi )
                                                        &=&     {\mathcal{B}} \mid \varphi  \mid ^2  +\kappa   \mid  \psi  \mid ^2
                                                                      - (\varphi \psi ^{\ast }  + \varphi ^{\ast }\psi  )   \\
       &=&
\begin{pmatrix}
  \varphi ^{\ast }   &   \psi ^{\ast }   \\    
\end{pmatrix}
{\mathcal{A}}^{\prime}
\begin{pmatrix}
     \varphi    \\   \psi    
\end{pmatrix},
\label{eqn:ssb28}     
\end{eqnarray}      
\end{subequations}      
with
\begin{equation}
{\mathcal{A}}^{\prime}:=
\begin{pmatrix}
  {\mathcal{B}}   &   -1   \\   -1   &  \kappa   
\end{pmatrix}
= ({{\mathcal{A}}^\prime})^{\dagger }.
\label{eqn:ssb29}   
\end{equation}
Also in this case,
it is clear that 
$   V_{U(1){\mathchar`-}{\rm{mini.}}}^{{\mathcal{J}}=0, \gamma  =1} $  takes quadratic form
and the matrix ${\mathcal{A}}^{\prime}$ is Hermitian.
Therefore  ${\mathcal{A}}^{\prime}$
can be diagonalized via an unitary matrix $ U^{\prime}$;
\begin{equation}
{ U^\prime}^{\dagger } {\mathcal{A}}^\prime U^\prime=
\begin{pmatrix}
  \lambda _{+}   &   0   \\   0   &  \lambda _{-}  
\end{pmatrix}.
\end{equation}
Each eigenvalue, $ \lambda _{\pm } $, is determined by the  following characteristic equation;
\begin{subequations}
\begin{eqnarray}
 \mid  \lambda  E  -  {\mathcal{A}}^{\prime}   \mid    &=&  0  
\label{eqn:ssb30}      
\end{eqnarray}            
\begin{eqnarray}
\Leftrightarrow 
  \lambda   &=&  \frac{  ( {\mathcal{B}}  + \kappa)  \pm   \sqrt{  ( {\mathcal{B}}  + \kappa)^2  -4 (  {\mathcal{B}}    \kappa -1)       }      }{2}   \label{eqn:ssb31-2}       \\
                     &=&  \frac{  ( {\mathcal{B}}  + \kappa)  \pm   \sqrt{  ( {\mathcal{B}}  - \kappa)^2  +4        }      }{2}    \label{eqn:ssb31-3}     \\
                     &=:& \lambda _{\pm },
\label{eqn:ssb31}      
\end{eqnarray} 
\end{subequations}           
with $  \lambda _{-}  <  \lambda_{+} $ by definition and note that $ 0< \lambda_{+} $.\footnote{
Remember that $0<  {\mathcal{B}} $  and  $  0< \kappa $.
}
According to eq. (\ref{eqn:ssb31-2}), 
$ \lambda _{-}$ becomes positive  when $1<  {\mathcal{B}}    \kappa  (\not=0) $; 
otherwise negative or zero ($   \lambda _{-} \leq 0$).

By using these eigenvalues $ \lambda _{\pm }$,
$   V_{U(1){\mathchar`-}{\rm{mini.}}}^{{\mathcal{J}}=0, \gamma  = 1} $
can be diagonalized as 
\begin{eqnarray}
    V_{U(1){\mathchar`-}{\rm{mini.}}}^{{\mathcal{J}}=0, \gamma = 1} ( \Phi^{\prime}_{+} ,   \Phi^{\prime}_{-} )&=& 
        \lambda _{+}   \mid  \Phi^{\prime}_{+}   \mid^{2}   + \lambda _{-}  \mid   \Phi^{\prime}_{-}   \mid^{2},   
\label{eqn:ssb34}        
\end{eqnarray}            
in which the   newly introduced complex scalar fields $  \Phi^{\prime}_{\pm }   $ are represented
by using an unitary matrix ${U^{\prime}}  $ as   
$  ( \varphi ^{\ast }      \    \    \psi ^{\ast }  )  {U^{\prime}}^{\dagger }  =: (  (\Phi^{\prime}_{+})^{\ast }    \   \    ( \Phi^{\prime}_{-}  )^{\ast }  ) $.
Fig. \ref{fig:SSBstability} (b) describes  $V_{U(1){\mathchar`-}{\rm{mini.}}}^{{\mathcal{J}}=0, \gamma  = 1}  $
when $  0  \leq  \lambda _{-}      $.
In this case,  though the origin (i.e. $  \mid  \Phi^{\prime} _+ \mid =\mid \Phi^{\prime} _-\mid =0  $)  is the stable vacuum state,
it is not generated by the $U(1)$-SSB;
it is simply the original vacuum
and it in fact gives $ \varphi _0 =0$. 
Let us remark that this can be easily confirmed also by the same procedure with sec. \ref{subsubsec:mini} (i.e. eq. (\ref{eqn:ssb15a}));
 $   V_{U(1){\mathchar`-}{\rm{mini.}}}^{{\mathcal{J}}=0, \gamma = 1}  (\varphi , \psi = \varphi/\kappa  )=  
({\mathcal{B}}-1/\kappa ) \mid  \varphi  \mid ^2   $.
On the other hand,
when $   \lambda _{-}  < 0 $, the situation is  the same with $   V_{U(1){\mathchar`-}{\rm{mini.}}}^{{\mathcal{J}}=\kappa ={\mathcal{B}}=0}  $;
Fig. \ref{fig:SSBstability} (a). 

Therefore, we conclude that 
the true stable vacuum state accompanied by the $U(1)$-SSB does not exist\cite{totsuka}
without the repulsive interaction;
$ {\mathcal{J}} \mid \varphi \mid ^4 $ with $0<  {\mathcal{J}} $.

\section{Theoretical proposal; \\
magnon BEC    in spin pumping system
}
\label{sec:magnonBEC}

We consider the application of the minimally generalized Goldstone model 
$  V_{U(1){\mathchar`-}{\rm{mini.}}}$
to  spin pumping systems;
the minimally generalized Goldstone model  can be regarded  
to describe the dynamics of magnons interacting with spins carried by conduction electrons at $ T=0$. 

\subsection{Anisotropic exchange interaction
}
\label{subsec:aniso}

In our previous work\cite{TSP} 
based on   the Schwinger-Keldysh formalism (i.e. a perturbative theory),
we have studied `thermal spin pumping'\cite{uchidaTSP} mediated by magnons 
in a ferromagnetic insulator and non-magnetic metal junction (Fig. \ref{fig:pumping}),
in which  localized spins $\mathbf{S} $  isotropically interact with  conduction electrons $ \mathbf{s} $   at the interface;
\begin{eqnarray}
  V_{\rm{iso.}} := -2 \gamma  ^{\prime}    \mathbf{S}   \cdot  \mathbf{s}  = -2 \gamma ^{\prime} (S^x s^x +S^ys^y+  S^zs^z).
\label{eqn:ssb001}
\end{eqnarray}
The variable $\gamma  ^{\prime}  (>0)$ represents the magnitude of the exchange interaction.
As far as our  `perturbative analysis',\cite{TSP}
the macroscopic condensate order parameter becomes zero
and magnon BEC cannot occur.
Now, by considering the correspondence of the spin pumping system  (i.e. magnon-electron system)
with the minimally generalized Goldstone model $V_{  U(1){\mathchar`-}{\rm{mini.}}}$,
we go after the possibility for the occurrence of the stable magnon BEC state 
in  spin pumping systems on the basis of a `non-perturbative theory'.

For the purpose, first,
let us consider the same situation\cite{TSP}
except the point that ferromagnetic localized spins $\mathbf{S} $  anisotropically   interact with  conduction electrons $ \mathbf{s} $; 
\begin{subequations}
\begin{eqnarray}
    V_{\rm{aniso.}} &:=&  -2 \gamma ^{\prime} (S^x s^x +S^ys^y+ \Delta S^zs^z)   \\
                                                              &=&   -2 \gamma ^{\prime} \Big( \frac{S^{+} s^{-} +S^{-}s^{+}}{2}  + \Delta S^zs^z   \Big),    
\label{eqn:ssb18}      
\end{eqnarray} 
\end{subequations}           
in which $ \Delta $ represents the magnitude of the anisotropic; $ 0\leq \Delta \leq 1 $.
Reflecting the fact that the minimally generalized Goldstone model (eq. (\ref{eqn:ssb12})) has not included  $ \mid \varphi \mid ^2 \mid \psi \mid ^2$,
we here focus on the strong anisotropic limit, $ \Delta \rightarrow 0$;
\begin{subequations}
\begin{eqnarray}
    V_{\rm{aniso.}}  &\stackrel{\Delta \rightarrow 0}{\longrightarrow } & - \gamma ^{\prime} ( S^{+} s^{-} +S^{-}s^{+}  )    \\
                                                      &=: & V_{\rm{aniso.}(\Delta =0)}.
\label{eqn:ssb19}  
\end{eqnarray}  
\end{subequations}          
Via the Holstein-Primakoff transformation;
$      S^{+}  = \sqrt{2S} a   +{\cal{O}}(1/{\sqrt{S}})  $,
 $     S^{-}  = \sqrt{2S} a^{\dagger }+{\cal{O}}(1/{\sqrt{S}})  $,
 $     S^z =S-a^{\dagger }a$,
 $V_{\rm{aniso.}(\Delta =0)}  $ can be expressed in terms of magnon creation/annihilation operators as follows;
\begin{eqnarray}
   V_{\rm{aniso.}(\Delta =0)} =  - \sqrt{2S} \gamma ^{\prime} ( a s^{-} + a^{\dagger } s^{+}  ).    
\label{eqn:ssb20}      
\end{eqnarray}            

We take the classical limit\cite{altland} and
each operator is replaced with a commutative complex scalar field (i.e. c-number);
    $a  \stackrel{\rm{classical}}{\mapsto  }  \varphi  \in  {\mathbb{C}} $,
    $s^{+}   \stackrel{\rm{classical}}{\mapsto  }  \psi   \in  {\mathbb{C}} $.
As the result, in the classical limit,
the strong anisotropic exchange interaction between magnons (i.e. spin waves) and conduction electrons $   V^{\rm{cla.}}_{\rm{aniso.}(\Delta =0)}$
is rewritten as follows;
\begin{eqnarray}
   V^{\rm{cla.}}_{\rm{aniso.}(\Delta =0)} =  - \sqrt{2S} \gamma ^{\prime}  ( \varphi  \psi ^{\ast } + \varphi ^{\ast } \psi  ).    
\label{eqn:ssb23}     
\end{eqnarray}            
It is clear that
this term corresponds to the $\gamma  $-term in the minimally generalized Goldstone model 
$  V_{U(1){\mathchar`-}{\rm{mini.}}}  = {\mathcal{B}} \mid \varphi  \mid ^2  + {\mathcal{J}}  \mid \varphi   \mid ^4   
                                                                           -\gamma  (\varphi \psi ^{\ast }  + \varphi ^{\ast }\psi  ) +\kappa   \mid  \psi  \mid ^2 $;
 in the language of the spin pumping system,
the ${\mathcal{B}}$-term describes the couplings with  the effective magnetic field along the z-axis for magnons, 
the $\kappa$-term represents the interaction between up-spins and down-spins of conduction electrons,
and 
 the ${\mathcal{J}}$-term corresponds to the magnon-magnon interaction.
Therefore, if these quantities satisfy the condition for the occurrence of  the $U(1)$-SSB shown in sec. \ref{subsubsec:mini}
(inequality  (\ref{eqn:ssb16}));
$   (0<) \  {\cal{B}} < (\gamma  ^2/{\kappa})  $,
the stable magnon BEC state can be realized even under the interaction with conduction electrons.
The vacuum expectation value $ \varphi _0$, which corresponds to the macroscopic condensate order parameter of magnons (i.e. spin waves),
becomes a non-zero value (see Fig. \ref{fig:SSBU1}); 
$   \mid \varphi _0\mid =\sqrt{  (\gamma  ^2/\kappa -{\mathcal{B}})/{{2{\mathcal{J}}}} }$.

Here let us again remark that,  as stressed in sec. \ref{subsubsec:mini},
for the occurrence of  the stable magnon BEC state in the spin pumping system,
the repulsive interaction between up-spins and down-spins of conduction electrons (i.e. $0<\kappa$) is essential
as well as the repulsive magnon-magnon interaction (i.e. $0<{\cal{J}}$),
which can be realized, as an example,\cite{totsuka}
owing to the dipolar interaction.\cite{tupitsyn}
This is the rigorous theoretical result based on `the non-perturbative theory'
beyond `a perturbative one'.\cite{TSP}

Last, it might be useful to mention that
in sec. \ref{subsubsec:stability},
we have closely investigated the stability of the vacuum   in the minimally generalized Goldstone model $ V_{U(1){\mathchar`-}{\rm{mini.}}}$.
There, we have concluded that
the true stable vacuum state accompanied by  the $U(1)$-SSB does not exist 
without the repulsive interaction (see Fig. \ref{fig:SSBstability});  $ {\mathcal{J}} \mid \varphi \mid ^4 $ with $0<  {\mathcal{J}} $.
This means, in  the language of the above spin pumping system, 
that  the true stable magnon BEC state cannot exist 
without the repulsive magnon-magnon interaction;
although the minimally generalized Goldstone model
$V_{U(1){\mathchar`-}{\rm{mini.}}}^{{\mathcal{J}}=0, \gamma  \equiv 1}  $ can possess the stable vacuum state 
when $  0  \leq  \lambda _{-}   $ (see Fig. \ref{fig:SSBstability} (b)),
it is not generated by the $U(1)$-SSB
and it in fact gives $\varphi _0=0$.
Therefore we suspect that magnon BEC cannot occur\cite{totsuka}
in the spin pumping system described by the minimally generalized Goldstone model with ${\cal{J}}=0$
(i.e. $V_{U(1){\mathchar`-}{\rm{mini.}}}^{{\mathcal{J}}=\kappa ={\mathcal{B}}=0}$  or $V_{U(1){\mathchar`-}{\rm{mini.}}}^{{\mathcal{J}}=0, \gamma  =1}$).
Here, 
let us point out that
BEC should not be identified with superfluid\cite{hickbec,altland,totsuka}
and hence there might  exist a superfluid phase in that case (i.e. ${\cal{J}}=0$).\cite{totsuka,bender}

\subsection{Generalized Goldstone model}
\label{subsec:geneGoldstone}

The next focus lies on whether the stable magnon BEC state could exist under a finite (i.e. non-zero) $\Delta $ regime
in  the above spin pumping system.\cite{TSP}
For the purpose,
we  include the term
 $     \mid \varphi  \mid ^2   \mid \psi    \mid ^2   $,
which arises from the $\Delta $-term in $V_{\rm{aniso.}}$
(eq. (\ref{eqn:ssb18}));
\begin{subequations}
\begin{eqnarray}
    V_{U(1)} (\varphi , \psi )  &:=&  V_{  U(1){\mathchar`-}{\rm{mini.}}} (\varphi , \psi )
                                                                -\alpha   \mid \varphi  \mid ^2   \mid \psi    \mid ^2    \\
                                                                &=&     {\mathcal{B}} \mid \varphi  \mid ^2  + {\mathcal{J}}  \mid \varphi   \mid ^4  
                                                                   -\alpha   \mid \varphi  \mid ^2   \mid  \psi    \mid ^2 \nonumber  \\
                                                                      &-&\gamma  (\varphi \psi ^{\ast }  + \varphi ^{\ast }\psi  ) +\kappa   \mid  \psi  \mid ^2,
\label{eqn:ssb36}       
\end{eqnarray}   
\end{subequations}         
where $  \alpha  (\in {\mathbb{R}})$ is the corresponding  dimensionless coupling constant.
From the viewpoint of  the correspondence with the above spin pumping system described by $V_{\rm{aniso.}}$,
we restrict  $\alpha $ to a positive value; $0<\alpha $.

In the language of the spin pumping system (i.e. the Holstein-Primakoff transformation),
the variable $ \mid \varphi \mid ^2 $ represents the number of magnons obeying the parastatistics\cite{kloss}
and hence, the relation;  $   \mid \varphi \mid ^2 < S={\cal{O}}(1)  $,  is required by definition.  
Moreover because we here treat the extremely low temperature regime (i.e. $ T=0$),
the variable $ \mid \varphi \mid ^2 $ is supposed to be very small 
enough to  satisfy the relation; $ \mid \varphi \mid ^2 \ll {\cal{O}}(1)  $.
Therefore
when we choose variables, $\kappa  $ and $ \alpha $,  to satisfy the condition; $ \kappa/\alpha ={\cal{O}} (1) $,
we are allowed to assume the relation;
$  \mid \varphi  \mid ^2 \ll  \kappa/\alpha    \Leftrightarrow  \alpha \mid \varphi  \mid ^2   \ll  \kappa  $.
Also from the viewpoint of the stability of the system in respect to $ \psi (\psi ^{*})$,
$  V_{U(1)} (\psi ) = (\kappa- \alpha \mid \varphi  \mid ^2) \mid \psi    \mid ^2  +  {\cal{O}}(\psi^{(\ast )} )    +  {\cal{O}} ({\psi^{(\ast) }}^{0} )   $,
the relation is strongly required.
Thus from now on,
we discuss on the basis of the assumption; $  \alpha \mid \varphi  \mid ^2   \ll  \kappa  $.
In other words,
the following our analysis is adequate in the region.

Through the same procedure with the minimally generalized Goldstone model $  V_{  U(1){\mathchar`-}{\rm{mini.}}}$
and the approximation; $ (\kappa - \alpha \mid \varphi \mid ^2)^{-1}\simeq (1+\alpha \mid \varphi \mid ^2/{\kappa})/{\kappa} $,
the  generalized Goldstone model $V_{U(1)} $  on the point,  
$ \psi = \gamma \varphi /(\kappa - \alpha  \mid \varphi \mid ^2)  \   \Leftarrow    \partial V_{U(1)}/ (\partial \psi ) =0$,
reads
\begin{subequations}
\begin{eqnarray}
V_{U(1)} (\varphi , \psi = \frac{\gamma }{\kappa - \alpha  \mid \varphi \mid ^2}  \varphi )
  & =& ({\mathcal{B}}  -  \frac{\gamma  ^2}{\kappa}) \chi    
                          + ({\mathcal{J}}  -  \frac{\alpha \gamma  ^2}{\kappa^2}) \chi ^2   
                           - \frac{2\alpha ^2\gamma  ^2}{\kappa^3} \chi ^3    \\
                           &=:&   V_{U(1)}(\chi ),
\label{eqn:ssb41}       
\end{eqnarray} 
\end{subequations}           
with  $  \chi := \mid \varphi \mid ^2  (>0) $.

\begin{figure}[h]
\begin{center}
\includegraphics[width=7cm,clip]{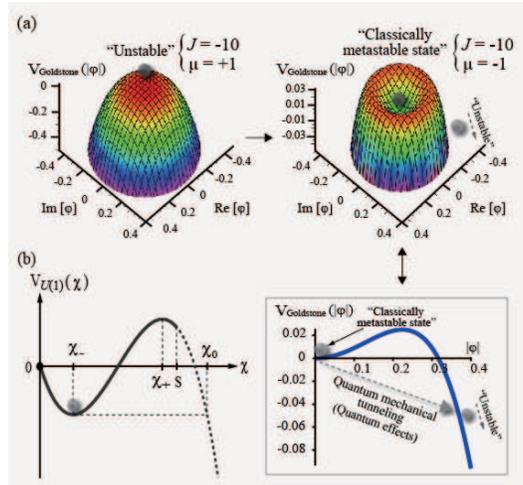}
\caption{(Color online).
(a) 
Plot of  the Goldstone model $ V_{\rm{Goldstone}}$ (eq. (\ref{eqn:ssb5})) with $ {\mathcal{J}}<0  $.
The $U(1)$-SSB of the vacuum does not occur.
The origin (i.e. $ \mid \varphi \mid =0$)  is unstable ($0\leq \mu$)  or  the classically metastable state ($\mu <0$).
(b) 
A schematic picture of the generalized Goldstone model $ V_{U(1)}(\chi )$ (eq. (\ref{eqn:ssb41})).
When $\chi _{-} \leq 0 $, the situation is the same with (a).
The state $  \chi _{0}$ is defined as 
$   V_{U(1)} (\chi= \chi _{0} ) = V_{U(1)} (\chi= \chi _{-} )$.
 \label{fig:SSBfeeble} }
\end{center}
\end{figure}

Here let us denotes the solution of the equation, $  d V_{U(1)}(\chi )/ (d\chi)  =0 $,
as $ \chi _{\pm }$  with  $  \chi _{-}  < \chi _{+}   $ by definition (see Fig.  \ref{fig:SSBfeeble} (b)).
The coefficient of $\chi ^3$ in $ V_{U(1)}(\chi ) $, $ 2\alpha ^2\gamma  ^2/(\kappa^3)  $, takes a positive value.
Therefore for the occurrence of the $U(1)$-SSB accompanied by the stable magnon BEC state, 
the condition is required (Fig. \ref{fig:SSBfeeble} (b));
\begin{subequations}
\begin{eqnarray}
    0 &<& \chi _{-}    \label{eqn:ssb43}      \\
    &\rm{and}&     \nonumber  \\
 S &<& \chi _{0}.  
\label{eqn:ssb43-2}       
\end{eqnarray}     
   \end{subequations}    
That is, when
\begin{subequations}
\begin{eqnarray}
      \frac{\alpha \gamma  ^2}{\kappa^2}  &<&   {\mathcal{J}},         \label{eqn:ssb45-2}                      \\
      \frac{\gamma  ^2}{\kappa}  -  \frac{\kappa^3}{6\alpha ^2\gamma  ^2}  
            \Big(  {\mathcal{J}}   -    \frac{\alpha \gamma  ^2}{\kappa^2}   \Big)^2  &<&  {\mathcal{B}}  <  \frac{\gamma  ^2}{\kappa},     \\
   &\rm{and}&   \nonumber   \\
   S &<& \chi _{0}
\label{eqn:ssb45}       
\end{eqnarray} 
\end{subequations}           
with
\begin{eqnarray}
    \chi _{0}
  &=&  \frac{\kappa^3}{6\alpha ^2\gamma  ^2}   \Bigg[
                 ({\mathcal{J}}  -\frac{\alpha \gamma  ^2}{\kappa^2})      
                  +2  \sqrt{    ({\mathcal{J}}  -\frac{\alpha \gamma  ^2}{\kappa^2})^2  
                  +  \frac{6\alpha ^2\gamma  ^2}{\kappa^3}   ({\mathcal{B}}  -  \frac{\gamma  ^2}{\kappa})   }    
            \Bigg],  
\label{eqn:ssb46-5}      
\end{eqnarray}        
the $ U(1)$-SSB accompanied by the stable magnon BEC state ($  \chi _{-} $)  occurs;
\begin{eqnarray}
    \chi _{-}
  &=&  \frac{\kappa^3}{6\alpha ^2\gamma  ^2}   \Bigg[
                 ({\mathcal{J}}  -\frac{\alpha \gamma  ^2}{\kappa^2})     
                  -  \sqrt{    ({\mathcal{J}}  -\frac{\alpha \gamma  ^2}{\kappa^2})^2  
                  +  \frac{6\alpha ^2\gamma  ^2}{\kappa^3}   ({\mathcal{B}}  -  \frac{\gamma  ^2}{\kappa})   }    \
            \Bigg].  
\label{eqn:ssb46}      
\end{eqnarray}       

Let us remark that
magnons obey the parastatistics and hence when $ S > \chi _{0}$,
the state $ \chi _{-}$ becomes the classically metastable  state\cite{peskin,coleman,wen}
and it does not give the absolute minimum.
That is, the state $ \chi _{-}$ is not the true stable vacuum and it can decay to the true vacuum by the quantum-mechanical tunneling effect\cite{peskin}
(see Fig. \ref{fig:SSBfeeble} (a) as an example).
Of course we have noted that we have been theoretically discussing within the classical theory,
but quantum effects are inevitable in real materials (i.e. experiments).
Thus, the condition  $ S < \chi _{0}$ (eq. (\ref{eqn:ssb45})) is required for the experimental realization of stable magnon BEC
in spin pumping systems with the non-zero $\Delta $-term (i.e. $\alpha $-term).

\section{Summary and discussion}
\label{sec:summary}

In order to go after the possibility for the stable magnon BEC state in  spin pumping systems,
we have employed a non-perturbative theory to go beyond  the perturbative analysis
and have extended the standard Goldstone model.
For the realization of the stable magnon BEC state,
the repulsive interaction  between up-spins and down-spins of conduction electrons  is essential
as well as the repulsive magnon-magnon interaction.
By realizing the condition we have clarified in sec. \ref{subsec:aniso}  and \ref{subsec:geneGoldstone} (it depends on materials),
the true stable magnon BEC state  can be experimentally observed also in spin pumping systems
without using external pumping magnetic fields. 

On the other hand,
to extend the system at finite temperature with quantum effects is left as a future work.
On top of this,
we consider that to clarify the effects of the unusual energy dispersion of the lowest magnon mode in YIG,
which is a relevant material to the experiment of magnon BEC\cite{demokritov,demidov,chumak} and spin pumping,\cite{spinwave,sandweg}
is a significant theoretical issue.
In addition, as stressed by Hick et al,\cite{hickbec}
BEC of quasiparticles is not necessarily accompanied by superfluidity.\cite{altland}
In other words, they should not be identified.\cite{totsuka}
Of course it is roughly expected, owing to Bogoliubov theory,\cite{altland} that superfluid of magnons is accompanied by magnon BEC in spin pumping systems,
but to reveal the detailed  relationship between magnon BEC and superfluid\cite{bunkovalone} of magnons in  spin pumping systems
is left as a important future work.

BEC state (i.e. coherent state) is the robust macroscopic quantum state against the loss of information.
Therefore we hope this work  becomes a bridge between the research on spintronics and magnon BEC
to lead to the green information technologies.

\section{Acknowledgements}
One of the authors (K. N.)  would like to thank 
K. Totsuka for stimulating the study and fruitful discussion about magnon BEC.
He is also grateful to M. Oshikawa, H. T. Ueda, Y. Bunkov, and G. Volovik
for helpful correspondences about magnon BEC.
He also would like to thank   the members of the Condensed Matter Theory and Quantum Computing Group of the University of Basel
for the warm hospitality and financial support 
during his stay under the young researchers exchange program by the Yukawa Institute for Theoretical Physics.
In particular, he is grateful to Dr. Kevin A. van Hoogdalem and Prof. Daniel Loss for significant discussion during his stay.

We are  supported by the Grant-in-Aid for the Global COE Program
"The Next Generation of Physics, Spun from Universality and Emergence"
from the Ministry of Education, Culture, Sports, Science and Technology (MEXT) of Japan.
One of the authors (Y. K.) is supported by the Grant-in-Aid for JSPS Fellows No. 24-4198.

\appendix
\section{Appendix:
Diagonalization of quadratic form}
\label{sec:diagonalization}

In this Appendix,
we  show the detail of the diagonalization in sec. \ref{subsubsec:stability}.
Remember  that 
any Hermitian matrices $ \mathcal{A}$  can be diagonalized via an unitary matrix $ U$ as follows;
\begin{equation}
 U^{\dagger } {\mathcal{A}}  U =
\begin{pmatrix}
  \lambda^{\prime} _{+}   &   0   \\      &     \lambda^{\prime} _{-} 
\end{pmatrix},
\label{eqn:ssb03}   
\end{equation}
where  $ \lambda^{\prime} _{\pm }$  represent eigenvalues  
which are determined by  the  following characteristic equation;  
$ \mid   \lambda^{\prime}  E -  {\mathcal{A}} \mid  = 0$.
Here the  ($ 2\times 2$) identity matrix is represented as $E$.
In addition,
the unitary matrix $U$ is constructed, 
via the eigenvector $ u_{\pm }$ which satisfy the relation; 
$ ( \lambda^{\prime} _{\pm }  E  -  {\mathcal{A}}) u_{\pm } =0$,
as $ U= ( u_{+}    \   \    u_{-}  ) $.

On the basis of the above procedure,
we  diagonalize $ V_{U(1){\mathchar`-}{\rm{mini.}}}^{{\mathcal{J}}=\kappa ={\mathcal{B}}=0} $
as an example;
\begin{subequations}
\begin{eqnarray}
    V_{U(1){\mathchar`-}{\rm{mini.}}}^{{\mathcal{J}}=\kappa ={\mathcal{B}}=0} (\varphi , \psi  )&=& 
    -\gamma  (\varphi \psi ^{\ast }  + \varphi ^{\ast }\psi  )   \\
 &=& -\gamma 
\begin{pmatrix}
  \varphi ^{\ast }   &   \psi ^{\ast }   \\    
\end{pmatrix}
{\mathcal{A}}
\begin{pmatrix}
     \varphi    \\   \psi    
\end{pmatrix},
\label{eqn:ssb04}   
\end{eqnarray}   
\end{subequations}         
with
\begin{equation}
{\mathcal{A}}:=
\begin{pmatrix}
  0   &   1   \\   1   &  0   
\end{pmatrix}
= {\mathcal{A}}^{\dagger }.
\label{eqn:ssb05}   
\end{equation}
It is clear that 
$  V_{U(1){\mathchar`-}{\rm{mini.}}}^{{\mathcal{J}}=\kappa ={\mathcal{B}}=0} $  takes quadratic form
and the matrix $\mathcal{A}$ is Hermitian.
Therefore  $  V_{U(1){\mathchar`-}{\rm{mini.}}}^{{\mathcal{J}}=\kappa ={\mathcal{B}}=0} $
can be diagonalized;
\begin{equation}
 U^{\dagger } {\mathcal{A}} U=
\begin{pmatrix}
  1   &   0   \\   0   &  -1  
\end{pmatrix},   \   \      \rm{with}    \    \
U :=  \frac{1}{\sqrt{2}}
\begin{pmatrix}
  1   &   1   \\   1   &  -1  
\end{pmatrix}.
\label{eqn:ssb06}   
\end{equation}
Note that the characteristic equation gives eigenvalues,
$ \lambda _{+}^{\prime} \equiv  +1  $ and $ \lambda _{-}^{\prime} \equiv  -1   $,
and the corresponding eigenvectors read
\begin{equation}
 u_{+} = \frac{1}{\sqrt{2}}
\begin{pmatrix}
       1   \\   1    
\end{pmatrix},    \     \   
u_{-} = \frac{1}{\sqrt{2}}
\begin{pmatrix}
       1   \\   -1   
\end{pmatrix}.
\label{eqn:ssb07}   
\end{equation}
As the result,
$  V_{U(1){\mathchar`-}{\rm{mini.}}}^{{\mathcal{J}}=\kappa ={\mathcal{B}}=0} $
can be rewritten as 
\begin{subequations}
\begin{eqnarray}
    V_{U(1){\mathchar`-}{\rm{mini.}}}^{{\mathcal{J}}=\kappa ={\mathcal{B}}=0} 
&=&
   -\gamma 
    \begin{pmatrix}
  \varphi ^{\ast }   &   \psi ^{\ast }   \\    
\end{pmatrix}
U U^{\dagger }  {\mathcal{A}} U  U^{\dagger }
\begin{pmatrix}
     \varphi    \\   \psi    
\end{pmatrix}           \\
  &=&  -\gamma  (  \mid  \Phi_{+}   \mid^{2}   -  \mid   \Phi_{-}   \mid^{2}   ),
\end{eqnarray}  
\end{subequations}          
where the newly introduced complex scalar fields $  \Phi_{\pm }   $ are represented
by using an unitary matrix $ U  $ as 
\begin{subequations}
\begin{eqnarray}
U^{\dagger }
\begin{pmatrix}
  \varphi    \\   \psi        
\end{pmatrix}
&=&  \frac{1}{\sqrt{2}}
\begin{pmatrix}
 \varphi + \psi     \\      \varphi    -   \psi        
\end{pmatrix}  \\
&=:& \begin{pmatrix}
 \Phi _+     \\     \Phi _-       
\end{pmatrix}. 
\label{eqn:ssb08}   
\end{eqnarray}       
\end{subequations}     

\bibliographystyle{unsrt}
\bibliography{PumpingRef}

\end{document}